\documentclass[aps,showpacs,11pt]{revtex4}
\usepackage{dcolumn}
\usepackage{graphicx}
\usepackage{amsmath}
\usepackage{amsfonts}
\usepackage{amssymb}
\usepackage{psfrag}
\usepackage{wrapfig}
\usepackage{subfigure}
\usepackage{makeidx}
\usepackage{bm}
\usepackage{epsf}
\usepackage{hyperref}
\usepackage{color}
\usepackage{multirow,dcolumn}
\usepackage{graphicx}
\def\be{\begin{equation}}
\def\ee{\end{equation}}
\def\ba{\begin{eqnarray}}
\def\ea{\end{eqnarray}}

\begin{document}

\title{Chaos in Born-Infeld-AdS black hole within extended phase space}

\author{Yong Chen, Haitang Li and Shao-Jun Zhang}
\email{2111709024@zjut.edu.cn;lhaitang@zjut.edu.cn;sjzhang84@hotmail.com}
\affiliation{Institute for Theoretical Physics $\&$ Cosmology, Zhejiang University of Technology, Hangzhou 310023, China}

\date{\today}

\begin{abstract}
	
	Born-Infeld-AdS black holes in extended phase space may possess phase structures resembling that of van der Waals fluid in four-dimensional spacetime. We study dynamics of its state,  which is in the unstable spinodal region initially on phase space, under time-periodic  thermal perturbations. By applying the Melnikov method, it is found that there exists a critical amplitude $\gamma_c$ of the perturbations, which depends on the Born-Infeld parameter $b$ and the black hole charge $Q$, such that chaos occurs for $\gamma > \gamma_c$. We found that larger $b$ or $Q$ makes the onset of chaos easier. Space-periodic thermal perturbations on its equilibrium state are also studied and there is always chaos for whatever the perturbation amplitude is.
\end{abstract}


\maketitle

\section{Introduction}

Since the remarkable discovery of Bekenstein and Hawking in 1970s~\cite{Bekenstein:1973ur,Hawking:1974sw}, black hole thermodynamics has attracted lots of attention and efforts in society of physicists. It opens a new window to the yet unsolved quantum gravity problem as well as providing a possible deep connection between black holes physics and other areas of fields via the AdS/CFT correspondence~\cite{Maldacena:1997re,Gubser:1998bc,Witten:1998qj}. Amounts of subsequent researches show that black holes (BHs) not only have temperature and entropy, they also possess rich phase structures and transitions resembling that of ordinary thermodynamic systems. Their phase structures become even much more richer in extended phase space, where the negative cosmological constant is considered as thermodynamic pressure. Within this framework, BHs behave in many respects similar to various chemical phenomena of ordinary thermodynamic systems, such as solid-liquid phase transition, liquid-gas transition of van der Waals fluid, triple points, reentrant phase transition (RPT), heat engines and etc, and thus is sometimes referred as BH chemistry~\cite{Kastor:2009wy,Kubiznak:2014zwa,Mann:2016trh}. In the extended phase space, the well-known Hawking-Page transition~\cite{Hawking:1982dh} between the thermal radiation and Schwarzschild-AdS black hole (SAdS BH) is analogous to the ordinary solid-liquid transition~\cite{Kubiznak:2014zwa}. Moreover, when SAdS BH is charged, it will exhibit a first-order large-black-hole/small-black-hole (LBH-SBH) transition in a canonical ensemble, which is in many ways analogous to the liquid-gas transition of van der Waals fluid~\cite{Chamblin:1999tk,Chamblin:1999hg,Dolan:2011xt,Kubiznak:2012wp}. In Refs.~\cite{Gunasekaran:2012dq,Altamirano:2013ane}, it is found that in certain range of temperature, for Kerr-AdS BH and Born-Infeld-AdS BH, a LBH-SBH-LBH phase transition appears  which is similar to the RPT observed in ordinary thermodynamic systems. There have been amounts of work on discussing various phase transitions for diverse BHs, please refer to the review Ref.~\cite{Kubiznak:2016qmn} and refs therein for more details.

As the similarity between BHs and ordinary thermodynamic systems, it is interesting to see if there are other phenomena present in ordinary thermodynamic systems also exist in BHs. In Ref.~\cite{Slemrod:1985}, the authors studied the dynamical behaviours of ordinary van der Waals fluid under temporal thermal perturbations. Initially, the system is quenched to the unstable spinodal region (see the definition below)  which allows homoclinic orbit in phase space. Under time-periodic thermal perturbations, with the Melnikov method~\cite{Melnikov:1963}, it is found that there exists a critical value $\gamma_c$ of the amplitude of the perturbations. And when $\gamma > \gamma_c$, the system will exhibit features of chaos and determination of its time evolution is impossible practically. Space-periodic thermal perturbation on equilibrium state is also considered, and is found that spatial chaos always exists for whatever the amplitude of the perturbation is. So, it is natural to see if such kinds of phenomena appear in BHs whose phase structures resembling that of van der Waals fluid. In Ref.~\cite{Chabab:2018lzf}, the authors studied this problem for Reissner-Nordstrom-AdS (RNAdS) BH, and found that similar chaotic features also exist. And the critical $\gamma_c$ for temporal chaos is found to depend on the BH charge $Q$ with more charge making the onset of chaos easier. Later, this study is extended to include the effect of higher curvature term, precisely the Gauss-Bonnet term, and similar chaotic features also found~\cite{Mahish:2019tgv}. Moreover, the critical $\gamma_c$ for temporal chaos depends on the BH charge $Q$ as well as the Gauss-Bonnet coupling constant $\alpha$. Larger $\alpha$ makes the onset of chaos easier, while the effect of $Q$ on $\gamma_c$ becomes complicated. 

Inspired by these work, it will be interesting to see if such chaotic feature exists for other kinds of BHs and if there is any universal law on the critical $\gamma_c$. So in this work we would like to extended the study to Born-Infeld-AdS (BIAdS) BH to see the effect of nonlinearity of electromagnetic field on the chaotic features. Thermodynamics and phase structures of BIAdS BH are rather rich which have been studied thoroughly in Refs~\cite{Gunasekaran:2012dq,Banerjee:2012zm,Zou:2013owa}. Depending on the value of $b Q$, where $b$ is the Born-Infeld parameter, BIAdS BH may exhibit van der Waals-like phase transition as well as RPT. We will concentrate on the former one. 

This work is organised as follows. In Sec. II, a brief review of the thermodynamics of BIAdS BH is given. In Sec. III, we consider temporal thermal perturbation as well as spatial thermal perturbation in two subsections respectively. The last section is summary and discussion.

\section{Thermodynamics of Born-Infeld-AdS black hole in extended phase space}

In this section, we will give a brief review of the thermodynamics of BIAdS BH in extended phase space.

The Einstein action with a Born-Infeld field in D-dimensional spacetime reads~\cite{Gunasekaran:2012dq,Banerjee:2012zm,Zou:2013owa}
\be{
\mathcal{I}=\frac{1}{16\pi} \int d^{D}x\sqrt{-g}[-2\Lambda+\mathcal{R}+\mathcal{L}(F)],
}\ee
where $\Lambda=-\frac{(D-1)(D-2)}{2\ell^{2}}$ is the negative cosmological constant with $\ell$ being the AdS radius. The nonlinear electromagnetic term $\mathcal{L}(F)$ is given by
\be{
\mathcal{L}(F)=4 b^2\left(1-\sqrt{1+\frac{F^{\mu\nu}F_{\mu\nu}}{2b^{2}}}\right),
}\ee
with $b$ being the Born-Infeld parameter characterising the nonlinearity of the electromagnetic field. When $b\rightarrow \infty$, one recovers the usual Maxwell term. It allows the following spherically symmetric charged black hole solution~\cite{Fernando:2003tz,Dey:2004yt,Cai:2004eh}
\begin{eqnarray}
ds^{2}&=&-f(r)dt^{2}+\frac{1}{f(r)}dr^{2}+r^{2}d\Omega^{2}_{D-2},\nonumber \\ F&=&dA,
\end{eqnarray}
with the function $f(r)$ and one-form $A$ given in terms of hypergeometric function as
\begin{eqnarray}
f(r) &=& 1+\frac{r^2}{l^2}-\frac{m}{r^{D-3}}+\frac{4b^2r^2}{(D-1)(D-2)}(1-\sqrt{1+\frac{(D-2)(D-3)q^2}{2b^2r^{2D-4}}})\nonumber \\&& +\frac{2(D-2)q^2}{(D-1)r^{2D-6}}\ _2{\cal F}_1 [\frac{D-3}{2D-4},\frac{1}{2},\frac{3D-7}{2D-4},-\frac{(D-2)(D-3)q^2}{2b^2r^{2D-4}}], \\ A&=&- \sqrt{\frac{D-2}{2(D-3)}}\frac{q}{r^{D-3}}\ _2{\cal {F}}_1 [\frac{D-3}{2D-4},\frac{1}{2},\frac{3D-7}{2D-4},-\frac{(D-2)(D-3)q^2}{2b^2r^{2D-4}}]dt.
\end{eqnarray}
Here $m$ and $q$ are related to the mass $M$ and charge $Q$ of the BIAdS BH as
\be{
Q=\frac{q\omega_{D-2}}{4\pi}\sqrt{\frac{(D-2)(D-3)}{2}},\ M=\frac{(D-2)\omega_{D-2}}{16\pi}m,
}\ee
where $\omega _{D-2}=\frac{2\pi^{\frac{D-1}{2}}}{\Gamma(\frac{D-1}{2})}$ is the volume of unit (D-2)-dimensional sphere.

In extended phase space, the cosmological constant is considered as the thermodynamic pressure with the precise relation as
\be\label{pressure}
P= - \frac{1}{8\pi} \Lambda = \frac{(D-1)(D-2)}{16\pi \ell^2}.
\ee
The Hawking temperature $T$ can be derived from the metric, in terms of which we can cast Eq.~(\ref{pressure}) into a form of equation of state as~\cite{Zou:2013owa}
\be{
P(v,T)=\frac{T}{v}+\frac{D-3}{\pi(D-2)v^2}-\frac{b^2}{4\pi}(1-\sqrt{1+\frac{\pi^{2}4^{2D-2}Q^2}{b^2\omega^2_{D-2}[(D-2)v]^{2D-4}}}),\label{StateEq}
}\ee
with $v \equiv \frac{4r_+}{D-2}$ thus identified as the specific volume and $r_+$ being the black hole horizon~\cite{Kubiznak:2012wp}.

The BIAdS BH has rather rich phase structures. It may exhibit the usual Hawking-Page transition or the RPT depending on the values of parameters $(b, Q)$ as well as the spacetime dimension $D$, which has been studied thoroughly in Refs.~\cite{Gunasekaran:2012dq,Banerjee:2012zm,Zou:2013owa}. For $D=4$ with $bQ>\frac{1}{2}$ or $D\geq 5$, the $P-v$ diagram is always found to be qualitatively similar to that of the van der Waals fluid, as depicted in Fig. 1. In fact, there exists a critical temperature $T_c$ under which a LBH-SBH phase transition may occur when increasing the pressure which resembles the phenomenon of gas-fluid phase transition. On the right panel of Fig. 1, we show an example of this transition. 
From the figure, we can see that the $P-v$ curve can be divided into three regions:
\begin{itemize}
\item $v \in [0,v_\alpha]$: Small BH region (corresponds to the fluid phase) where $\frac{\partial P(v,T_0)}{\partial v}<0$;
\item $v \in [v_\alpha,v_\beta]$: Unstable region, called spinodal region, where $\frac{\partial P(v,T_0)}{\partial v}>0$. The two extreme points $v_\alpha$ and $v_\beta$ are determined by $\frac{\partial P(v,T_0)}{\partial v}\mid_{v=v_\alpha}=\frac{\partial P(v,T_0)}{\partial v}\mid_{v=v_\beta}=0$. There is also an inflection point $v=v_0$ with $\frac{\partial^2 P}{\partial^2 v}=0$;
\item $v \in [v_\beta, \infty]$: Large BH region (corresponds to the gas phase) where $\frac{\partial P(v,T_0)}{\partial v}<0$.
\end{itemize}

\begin{figure}[!htbp]
	\includegraphics[width=0.45\textwidth]{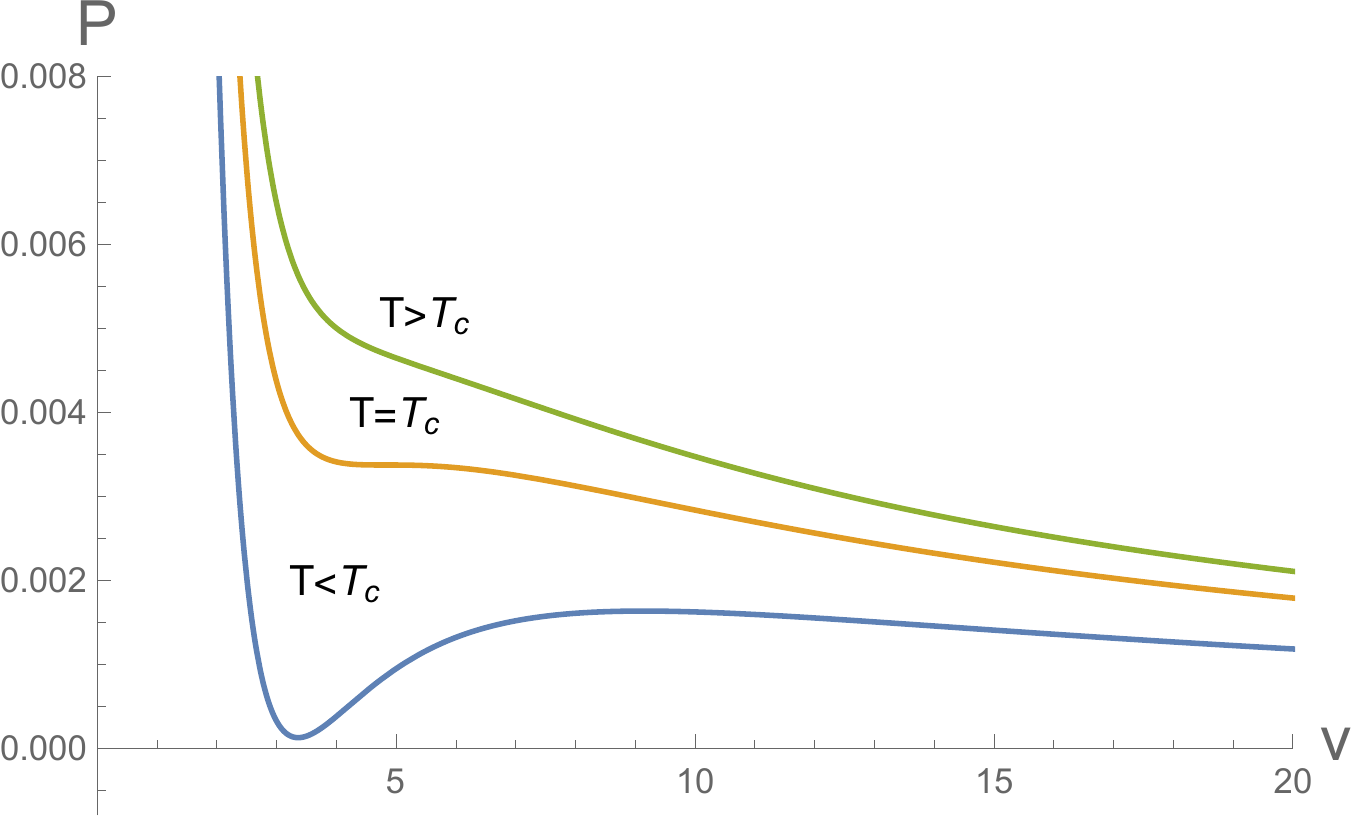}\quad
	 \includegraphics[width=0.45\textwidth]{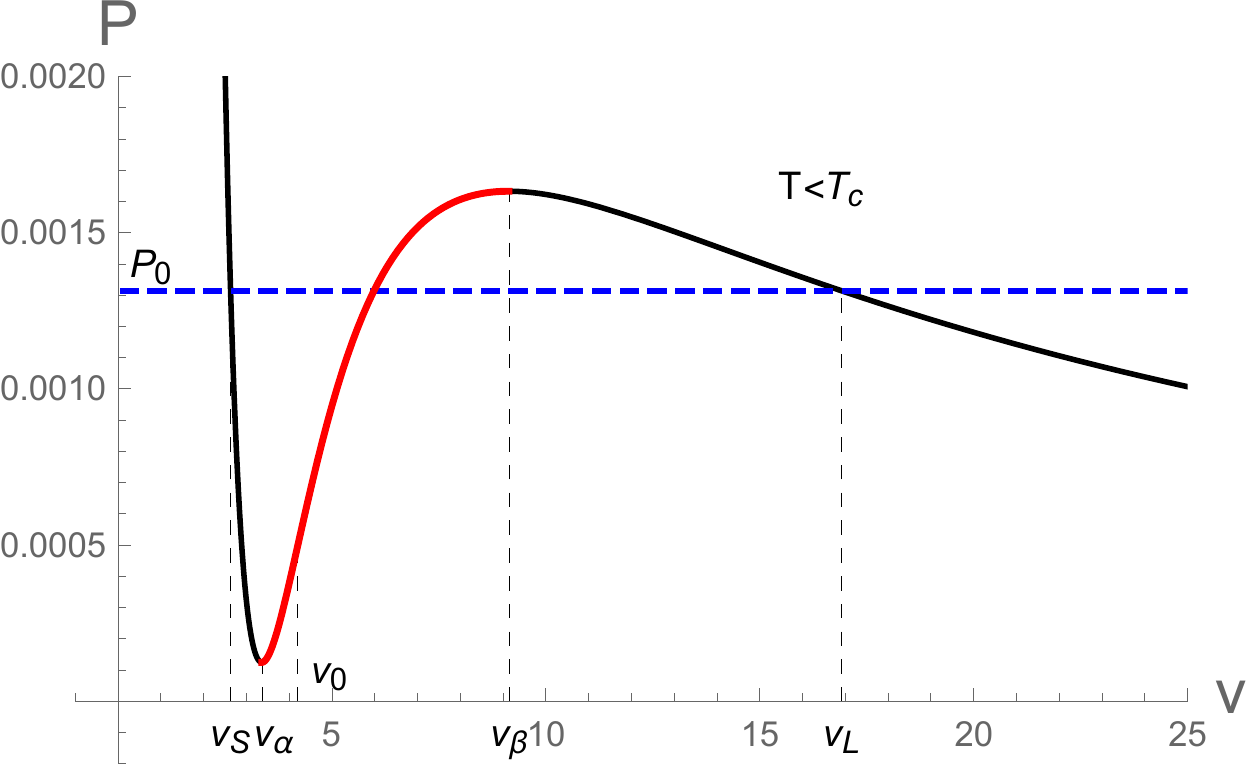}
	\caption{(colour online) The $P-v$ diagram of the BIA BH for different $T$. The parameters are fixed as $b=1, Q=1$ and $D=4$. On the right panel with $T< T_c$, the curve can be divided into three regions, two are stable (black ones) and one is unstable (red one). The blue dashed line denotes the coexisting line of large BH (with specific volume $v_L$) and small BH (with specific volume $v_S$) with transition pressure $P_0$, which can be determined by the "Maxwell area law".}
\end{figure}

In this paper, we are going to study the behaviours of the BIAdS BH under periodic thermal perturbations. We will show later that, under certain conditions, chaos appears. Following Ref.~\cite{Slemrod:1985}, we will consider two types of thermal perturbations and their corresponding chaos:
\begin{itemize}
	\item Temporal chaos in the spinodal region. In this case, the BH is initially in the unstable spinodal region. This state can be achieved by reducing the temperature of the fluid instantaneously starting from a stable state above $T_c$. Especially, we will consider the BH to be at $v=v_0$ in the spinodal region (see the right panel of Fig. 1). And then we will examine the effect of a time-periodic thermal perturbation. 
	
	\item Spatial chaos in the equilibrium state. In this case, the BH is initially in an equilibrium state and a space-periodic thermal perturbation is considered.
\end{itemize}
For simplicity, we will concentrate on the four-dimensional case. Extension to higher dimensions is straightforward.

\section{Chaos under thermal perturbations}

As Ref.~\cite{Slemrod:1985}, we assume that the flow of the BH fluid is planar and along  $x$-axis in a tube of unit cross section of fixed volume. One can then use one Eulerian coordinate $x$ to denote the fluid particle within the tube of flow. Let $x_\ast$ be the coordinate of the reference fluid particle. Then the mass $M$ of a column of fluid of unit cross section between the reference fluid particle and a fluid particle with Eulerian coordinate $x$ can be expressed as
\be{
M=\int^{x}_{x_\ast}\rho(\xi,t)d\xi,}
\ee
here $\rho(x,t)$ is the fluid density at position $x$ and time $t$. From this relation, the position of any particle can instead be defined by the mass as $x=x(M,t)$. Furthermore, from this relation, it is easy to get $x_M(M,t)\rho(x(M,t),t)=1$, where subscript means partial derivative, i.e., $x_M(M,t) \equiv \frac{\partial x}{\partial M}$. So, we have $x_M(M,t)= \rho(x(M,t),t)^{-1}$ which is just the specific volume $v(M,t)$. We can also define the velocity as $u(M,t) \equiv x_t (M,t)$.

\subsection{Melnikov method}

The Melnikov method, applied in this work, is one of the primary tools to judge the existence of chaos in a perturbed Hamiltonian system. Generally, dynamical equation of the system in phase space takes a form as
\be
\dot {\mathbf{z}} = f (\mathbf{z}) + \epsilon g(\mathbf{z},t), \label{GeneralEq}
\ee 
where the column vector $\bf{z}$ denotes the dynamical variables in phase space which has two degrees of freedom in our considered cases. The dot means derivative to the orbit parameter $t$ which may be the time or else. $\epsilon$ is a small perturbation parameter. 

When $\epsilon=0$, it is supposed that the dynamical equation allows a solution
\be
\mathbf{z} = \mathbf{z}_0 (t),
\ee 
which may be a homoclinic orbit  connecting a saddle point in phase space to itself, or a heteroclinic orbit connecting two saddle points. In the presence of perturbation $\epsilon \neq 0$, the homoclinic orbit or heteroclinic orbit may break to lead possible chaos whose existence can be determined by the Melnikov function. In our considered cases, the Melnikov function takes the following form
\be{
	M(t_0)=\int_{-\infty}^ \infty f(\mathbf{z}_0(t-t_0))\mathbf{J}  g(\mathbf{z}_0(t-t_0))dt, \quad \textrm{with}\ \mathbf{J}=\left [ \begin{array}{cc} 0 &1 \\-1 & 0 \end{array} \right ].\label{MelnikovFunc}
}\ee
If $M(t_0)$ crosses zero, chaos exists. For more details on Melnikov method and its application, please refer to Refs.~\cite{Holmes:1990,Aslanov:2017} and refs therein.

\subsection{Temporal chaos in spinodal region}

In this subsection, we will focus on the first type of chaos. The system is at $v=v_0$ with temperature $T=T_0<T_c$.  Then we consider a time-periodic thermal perturbation,
\be{
T=T_0 +\epsilon \gamma \cos{\omega t}\cos M,
}\ee
where $\gamma$ is amplitude of the perturbation relative to the small viscosity, as we will see more clearly later. From the balance of mass and momentum, we have the following equations
\begin{eqnarray}
 \frac{\partial v}{\partial t} &=& \frac{\partial u}{\partial M}, \\ \frac{\partial u}{\partial t} &=& \frac{\partial \tau}{\partial M},\label{MomentumBalance}
\end{eqnarray}
 where $\tau$ is the Piola stress. By assuming that the fluid is thermoelastic, slightly viscous, isotropic, the Piola stress takes a form as
 \be{
 \tau = -P(v,T)+\mu u_M-A v_{MM},
 }\ee
according to the Van der walls-Korteweg theory~\cite{Felderhof:1970}. Here $A$ is a positive constant, and $\mu$ is an assumed positive and small constant viscosity. With the stress form, the balance equation (\ref{MomentumBalance}) becomes
 \be{
 x_{tt}=-P(v,T)_M+\mu u_{MM}-Av_{MMM},
  }\ee
 Suppose that the total mass of the BH fluid is $2\pi/ s$ with $s$ a positive parameter. As the viscosity is assumed to be small, we rewrite it as $\mu \equiv \epsilon \mu_0$ with $\mu_0$ a constant. Then one can see that the amplitude ratio between the perturbation and viscosity is $\gamma/\mu_0$, so $\gamma$ can be seen as the relative amplitude of the perturbation as we stated above. For later convenience, we introduce the change of variables $\tilde {M}=s M, \tilde{t}=s t, \tilde{x}=s x$. Then, the mass range becomes $\tilde{M} \in [0, 2\pi]$ and the above equation can be written as (with overbars omitted)
  \be{
 x_{tt}=-P(v,T)_M+\epsilon \mu_0 s u_{MM}-A s^2v_{MMM}.\label{MainEq} 
  }\ee

The dynamical equation (\ref{MainEq}) can be solved by perturbation method. To do so, we expand $P(v,T)$ around the equilibrium point $(v_0,T_0)$
\begin{eqnarray}
P(v,T)&=&P(v_0,T_0)+P_v(v_0,T_0)(v-v_0)+P_T(v_0,T_0)(T-T_0)+P_{vT}(v_0,T_0)(v-v_0)(T-T_0) \nonumber \\&& +\frac{P_{vvv}(v_0,T_0)}{3!}(v-v_0)^3+\frac{P_{vvT}(v_0,T_0)}{2}(v-v_0)^2(T-T_0) +\cdots,\label{PExpansion}
\end{eqnarray}
with $P_{TT}(v_0,T_0)=P_{vTT}(v_0,T_0)=P_{TTT}(v_0,T_0)=0$ resulting from (\ref{StateEq}). As $v=v_0$ is the inflection point in the spinodal region, we have $P_{vv}(v_0,T_0)=0$ and $P_v(v_0,T_0)>0,P_{vvv}(v_0,T_0)<0$. Meanwhile, we expand $v, u$ in Fourier sine and cosine series respectively, on $M\in[0,2\pi]$, to obtain
\begin{eqnarray}
v(M,t)&=&x_M(M,t)=v_0+x_1 (t)\cos M + x_2 (t) \cos 2M + x_3 (t) \cos 3M + \cdots,\nonumber \\
u(x,t)&=&x_t(M,t)=u_1(t)\sin M + u_2 (t) \sin 2M + u_3 (t) \sin 3M + \cdots.\label{ModesExpansion}
\end{eqnarray}
Here $x_i (t)$ and $u_i(t)$ are considered to be hydrodynamical modes which describes the deviation from the original static equilibrium state with $v=v_0$. For simplicity, we only consider the first mode and neglect the effect of the higher modes (In fact, even if we consider the second mode as done in Refs.~\cite{Slemrod:1985,Chabab:2018lzf,Mahish:2019tgv}, it has no effect on our following discussions.). Substituting the expansions Eqs. (\ref{PExpansion}) (\ref{ModesExpansion}) into the dynamical equation (\ref{MainEq}), we can find~(omitting $(v_0,T_0)$)
\begin{eqnarray}
\dot{x}&=&u, \nonumber \\
\dot{u}&=&(P_v-A s^2)x+\epsilon \left(P_T+\frac{3P_{vvT}}{8}x^2 \right)\gamma \cos{\omega t}+\frac{P_{vvv}}{8}x^3-\epsilon \mu_0 s u,
\end{eqnarray}
where the subscript "1" in $x_1$ and $u_1$ has been omitted. The corresponding Hamiltionian can be constructed from the dynamical equations and takes the same form as in Ref.~\cite{Slemrod:1985}. By denoting $\mathbf{z} \equiv [x, u]^T$, the above equations can be cast into the form as Eq.~(\ref{GeneralEq}) with
\be{
	f(z)=
	\left [
	\begin{array}{c}
		u \\
		a^2 x +\frac{P_{vvv}}{8} x^3
	\end{array}
	\right ], \label{fFunction}
}\ee
and
\be{
	g(z)=
	\left [
	\begin{array}{c}
		0 \\
		\left( P_T+\frac{3 P_{vvT}}{8} x^2 \right)\gamma \cos {\omega t}-\mu_0 s u
	\end{array}
	\right ],\label{gFunction}
}\ee
where $a^2 \equiv (P_v-A s^2)$.

To apply the Melnikov method, firstly we need to solve the unperturbed equation $
\dot{\bf{z}}=f(\bf{z}).$ Luckily, it allows an analytical solution (with two branches) as~\cite{Holmes:1979}
\be{
	\bf{z}_0(t)=
	\left [
	\begin{array}{c}
		x_0(t) \\
		u_0(t)
	\end{array}
	\right ]=
	\left [
	\begin{array}{c}
		\frac{ \pm 4a}{(-P_{vvv})^{1/2}} \text{sech} (at) \\
		\frac{\mp 4a^2}{(-P_{vvv})^{1/2}} \text{sech} (at) \tanh (at)
	\end{array}
	\right ],} \label{HomoclinicOrbit}
\ee
which describes a homoclinic orbit connecting the origin to itself, as shown in Fig. 2. The two wings of this butterfly-like orbit correspond to the two branches of the solution (\ref{HomoclinicOrbit}) respectively. As $t \rightarrow \pm \infty$, $\bf{z}_0$ approaches to the origin point which is a saddle point.

\begin{figure}[!htbp]
	\centering
 \includegraphics[width=0.7\textwidth]{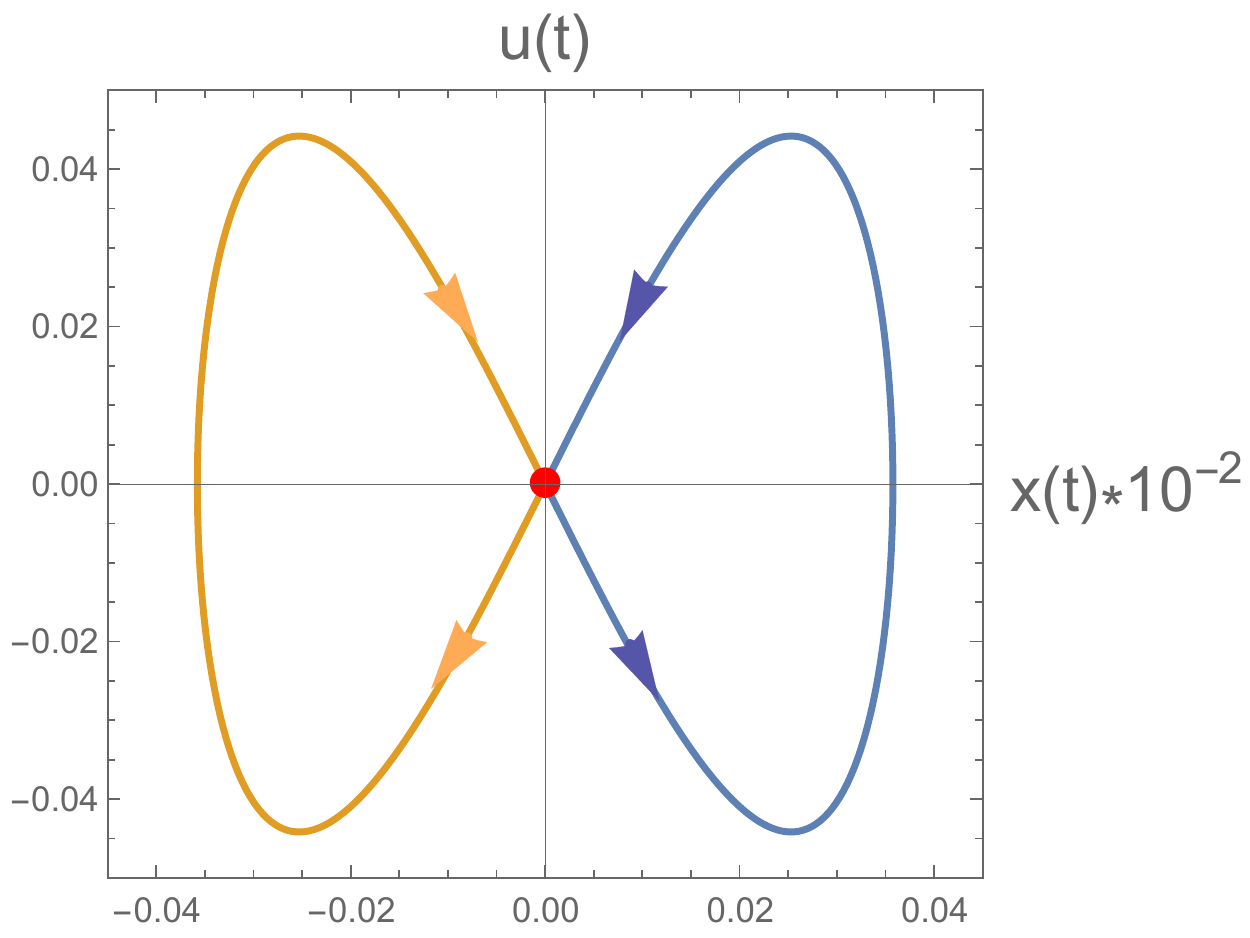}
\caption{(colour online) Homoclinic orbit of the unperturbed equations with $T=0.0315<T_c$. The parameters are fixed as $D=4, Q=1, b=1, A=0.2, s=0.001$. The two wings of this butterfly-like orbit correspond to the two branches of the solution respectively. The arrows denote the time flow direction. The origin point is marked with red dot.}
\end{figure}

Then we study the effect of perturbation ($\epsilon \neq 0$). 
With Eqs. (\ref{MelnikovFunc}) and (\ref{fFunction})(\ref{gFunction}), one can obtain
\begin{eqnarray}
M(t_0)&=& \int_{-\infty}^\infty dt \frac{4a^2}{(-P_{vvv})^{1/2}}\gamma \cos \omega t \left[\frac{6a^2P_{vvT}}{P_{vvv}} \text{sech}^2 a(t-t_0)-P_T \right]\text{sech} a(t-t_0) \tanh a(t-t_0) \nonumber \\&&+\int_{-\infty}^\infty dt \frac{16\mu_0 s a^4}{P_{vvv}} \text{sech}^2 a(t-t_0) \tanh^2 a(t-t_0).
\end{eqnarray}
The above integral can be finished with the residue theorem, and then we get
\be{
M(t_0)=\gamma \omega K sin \omega t_0+\mu_0 s L,
}\ee
with
\be{
K=\frac{8 \pi}{(-P_{vvv})^{1/2}} \left[P_T-\frac{P_{vvT}}{P_{vvv}}(\omega^2+a^2) \right]\frac{\exp(\frac{\pi \omega}{2a})}{1+\exp(\frac{\pi \omega}{a})},\qquad
L=\frac{32a^3}{3P_{vvv}}.
}\ee
Then, it is easy to see that $M(t_0)$ will have a simple zero if
\be{
\left|\frac{s\mu_0L}{\gamma \omega K}\right|\leq1.
}\ee
This relation defines a critical $\gamma_c$ which depends on various other parameters. And there will be temporal chaos when $\gamma > \gamma_c$. 

\begin{figure}[!htbp]
	\centering
	\subfigure[~~$\gamma = 0.000015<\gamma_c$]{\includegraphics[width=0.48\textwidth]{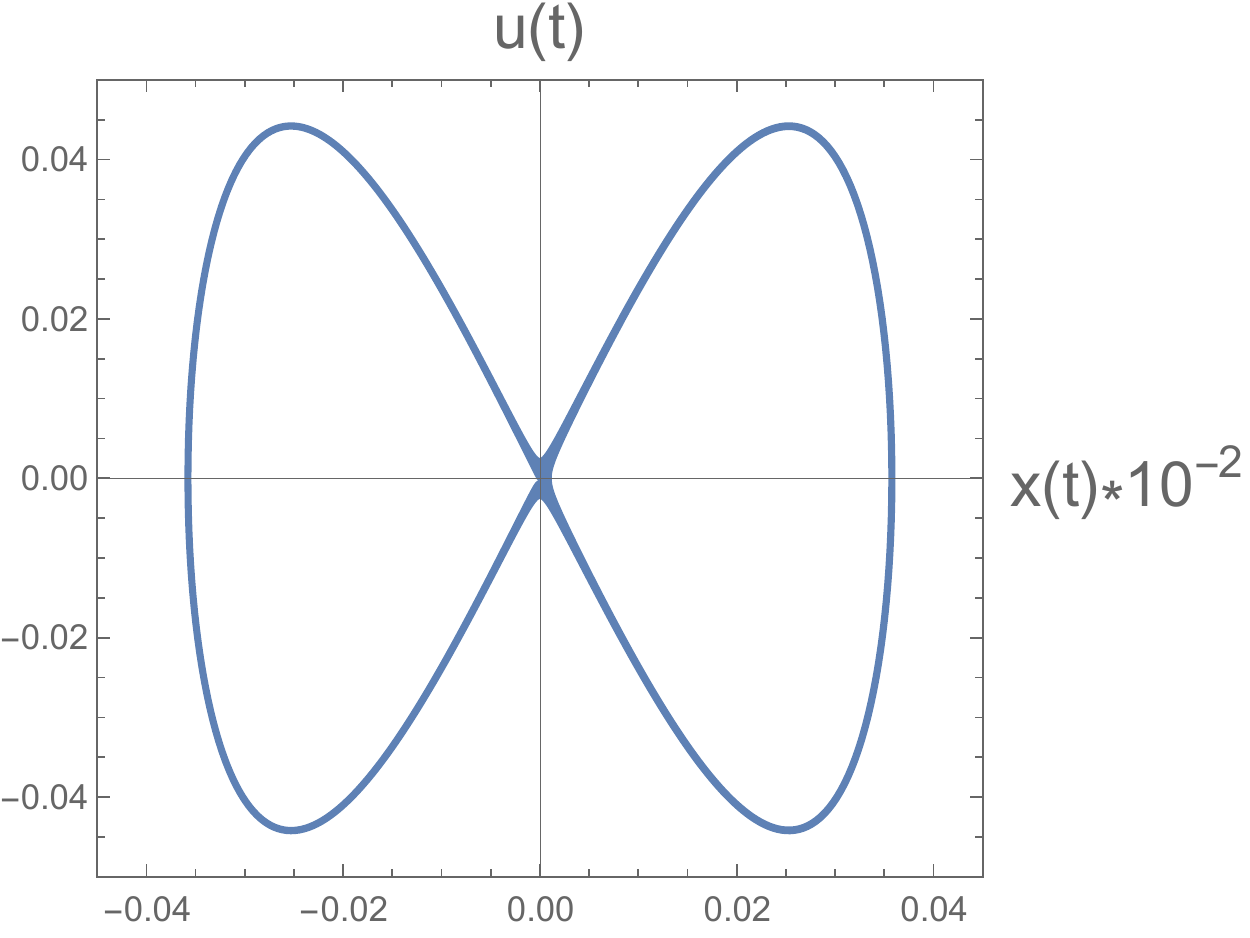}}
	\subfigure[~~$\gamma =0.5>\gamma_c$]{\includegraphics[width=0.48\textwidth]{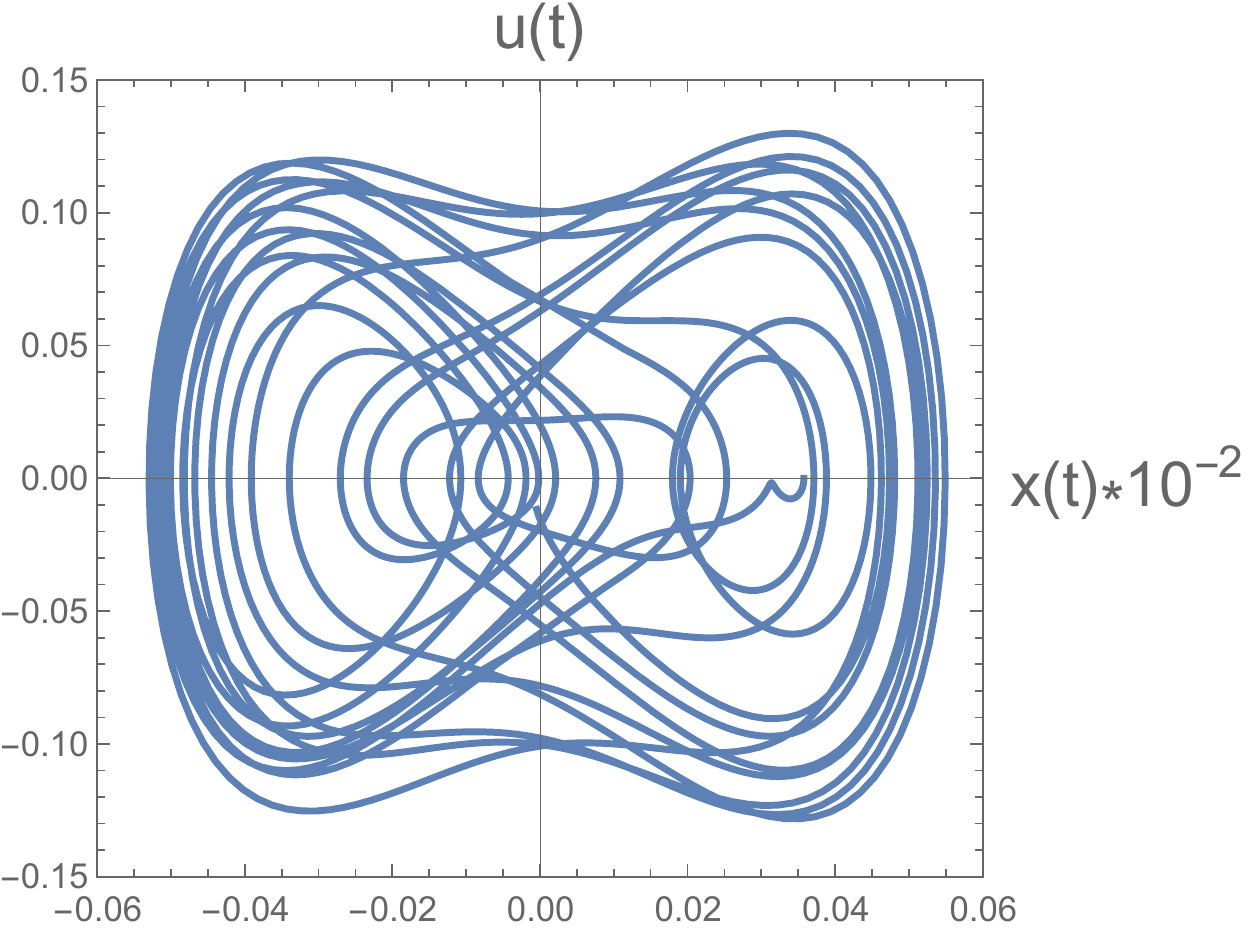}}
	\caption{(colour online) Time evolution of the perturbed equations in $x-u$ plane with $T=0.0315<T_c$. Parameters are fixed as $\omega=0.01, \epsilon=0.001, \mu_0 = 0.1 $ with other parameters set as in Fig. 2. The critical value $\gamma_c \approx 0.0000213$.}
\end{figure}
We solve the perturbed equation numerically. In Fig. 3, the time evolution of the state of the BH fluid in $x-u$ plane is plotted. From the figure, we can see that when $\gamma<\gamma_c$, the perturbation decays with time and the fluid will finally approach to its original equilibrium state with $v=v_0$ (although unstable); While for $\gamma>\gamma_c$, the fluid exhibits chaotic feature.

The critical value $\gamma_c$ depends on various other parameters, as we stated above. In Fig. 4, we show the dependence of $\gamma_c$ on the Born-Infeld parameter $b$ or the charge $Q$, with other parameters fixed. From the figure, one can see that larger $b$ or $Q$ makes the onset of chaos easier.

\begin{figure}[!htbp]
  \subfigure[~~$Q=1$]{\includegraphics[width=0.45\textwidth]{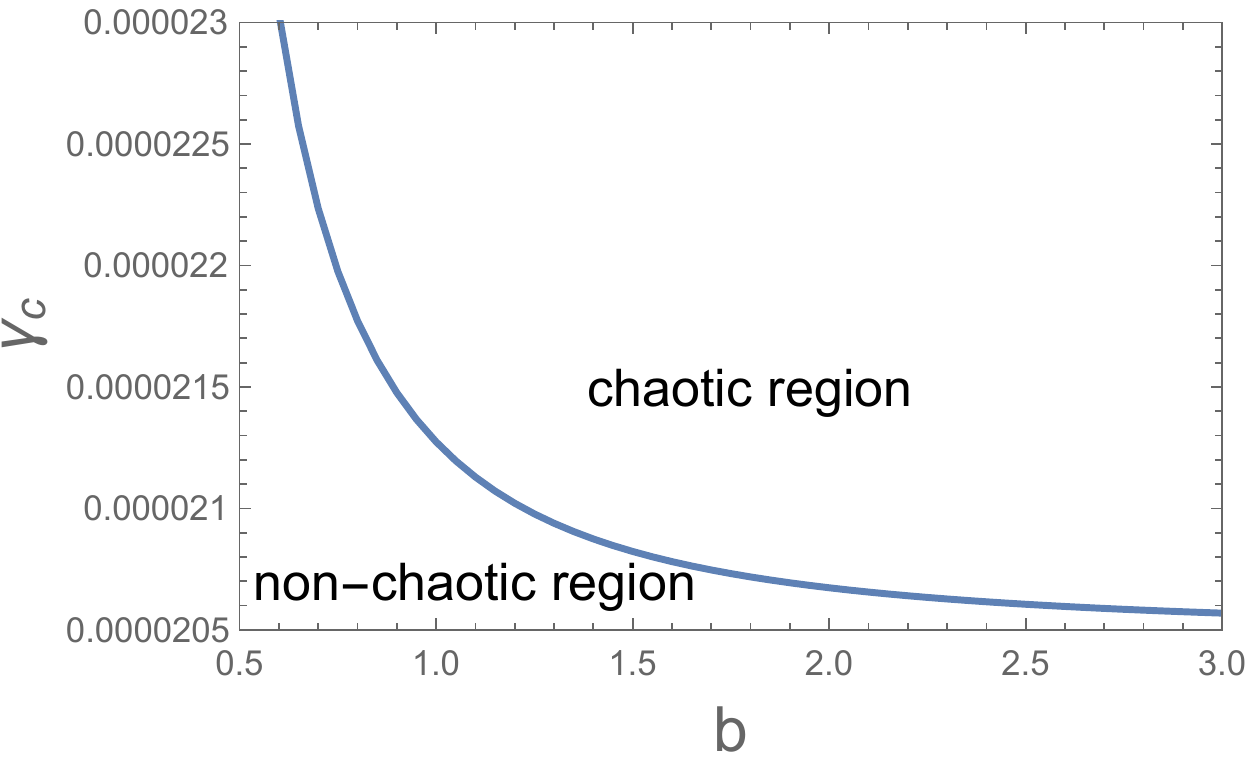}}\quad
  \subfigure[~~$b=1$]{\includegraphics[width=0.42\textwidth]{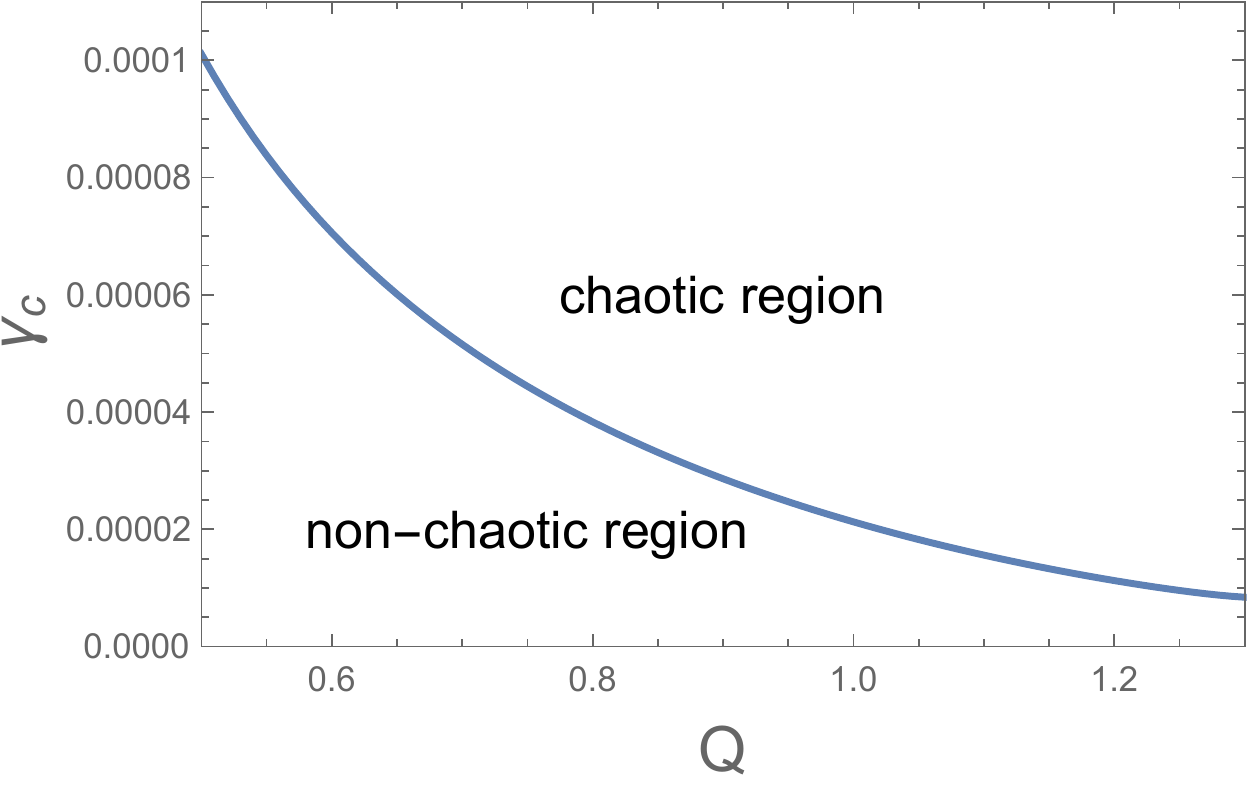}}
  \caption{(colour online) Critical values $\gamma_c$ versus $b$ or $Q$. Other parameters are fixed as in Fig. 3.}
\end{figure}

\subsection{Spatial chaos in the equilibrium state}

Now let us consider the effect of spatially periodic thermal perturbations on the equilibrium state of the BIAdS BH at temperature $T_0<T_c$. Without flow, the stress takes the following form according to the van der waals-Korteweg theory~\cite{Felderhof:1970}
\be{
\tau =-P(v,T_0)-A v^{''},\label{Stress}
}\ee
where $P$ is again given in Eq. (\ref{StateEq}), $A>0$ is a positive constant, and $'$ denotes $d/dx$. In the absence of body forces the balance of linear momentum is $\tau'=0$ which yields $\tau =B$ ($B$ is the constant stress at $|x|=\infty$, representing the ambient pressure). Then Eq. (\ref{Stress}) becomes
\be{
A v''+P(v,T_0)=B, \quad -\infty<x<\infty. \label{MainEq2}
}\ee
The portrait of $P(v,T_0)$ has been depicted in Fig. 1, with the transition point $v_S, v_L$ determined by the Maxwell's equal area law,
\be{
\int^{v_L}_{v_S} [P(v,T_0)-P(v_S,T_0)]dv=0,
}\ee

According to the value of the constant ambient pressure $B$, Eq. (\ref{MainEq2}) will yield three different types of portraits in the $v-v'$ phase plane:
\begin{itemize}
\item {\em Case 1}: $P_0<B<P(v_{\beta},T_0)$.

 The values $v_1, v_2, v_3$ so that $P(v_1,T_0)=P(v_2,T_0)=P(v_3,T_0)=B$ are shown in Fig. 5(a), while the portrait of Eq. (\ref{MainEq2}) in $v-v'$ phase plane is shown in Fig. 5(b);
 
\item {\em Case 2}: $P(v_{\alpha},T_0)<B<P_0$. 

The portrait of Eq. (\ref{MainEq2}) in $v-v'$ phase plane is shown in Fig. 6(b);

\item {\em Case 3}: $B=P_0$. 

The portrait of Eq. (\ref{MainEq2}) in $v-v'$ phase plane is shown in Fig. 7(b);
\end{itemize}
\begin{figure}[!htbp]
\centering
\subfigure [~~]{\includegraphics[width=0.48\textwidth]{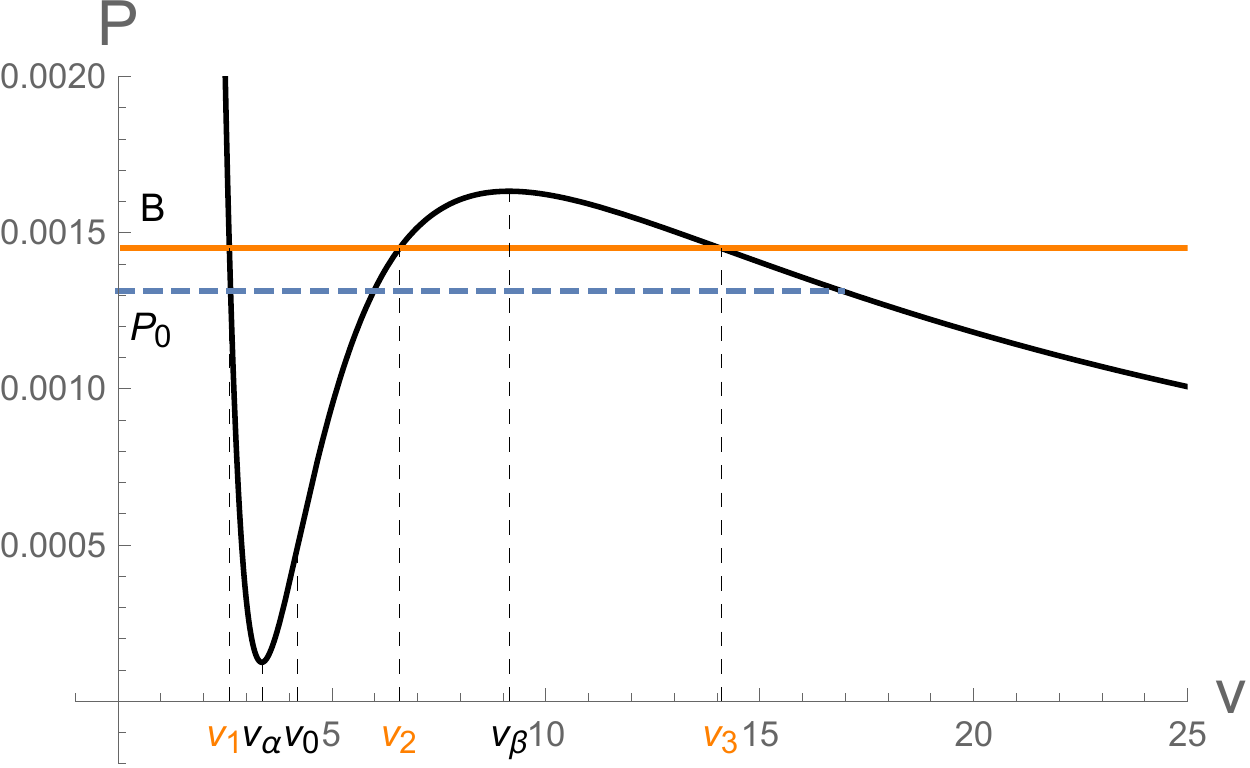}}\quad
\subfigure [~~]{\includegraphics[width=0.48\textwidth]{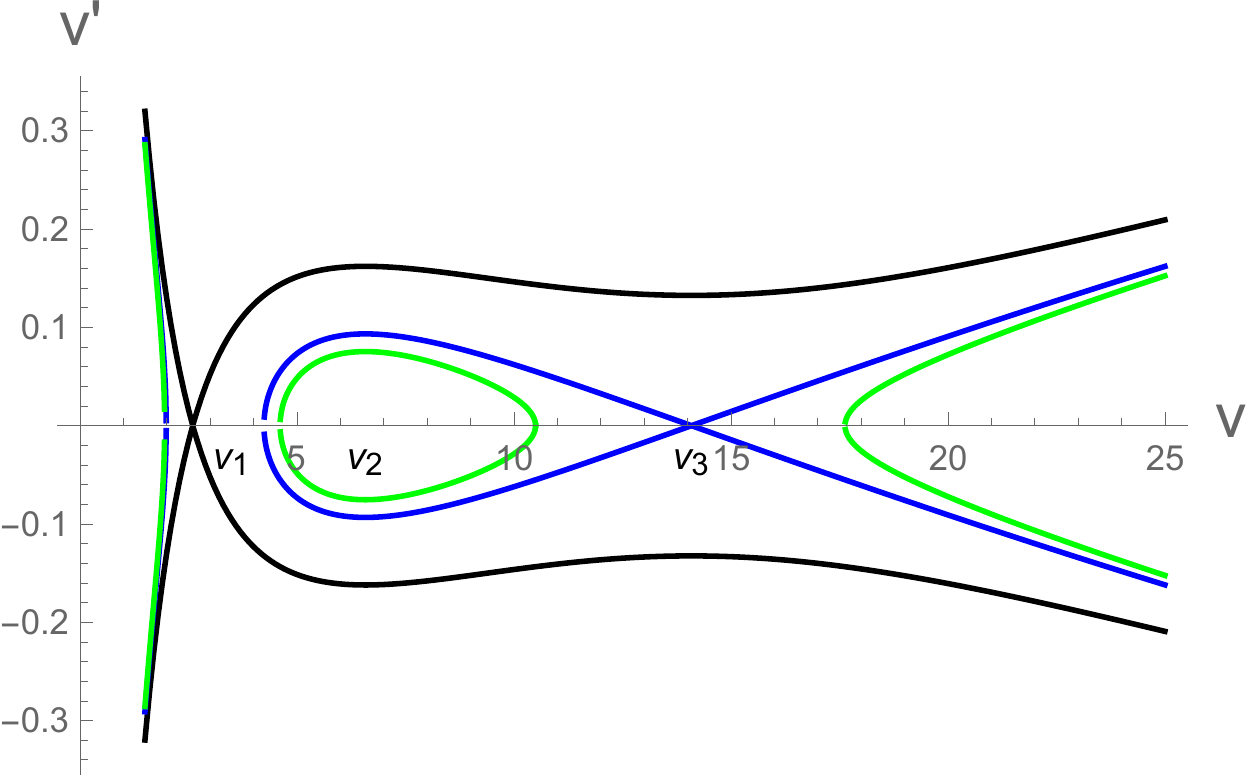}}
\caption{(colour online) Case 1: (a) $B$ and $v_1,v_2,v_3$ in $P-v$ plane; (b) $v-v'$ phase portrait. There is a homoclinic orbit connecting $v_3$ to itself (blue one).}
\end{figure}
\begin{figure}[!htbp]
\centering
\subfigure[]{\includegraphics[width=0.48\textwidth]{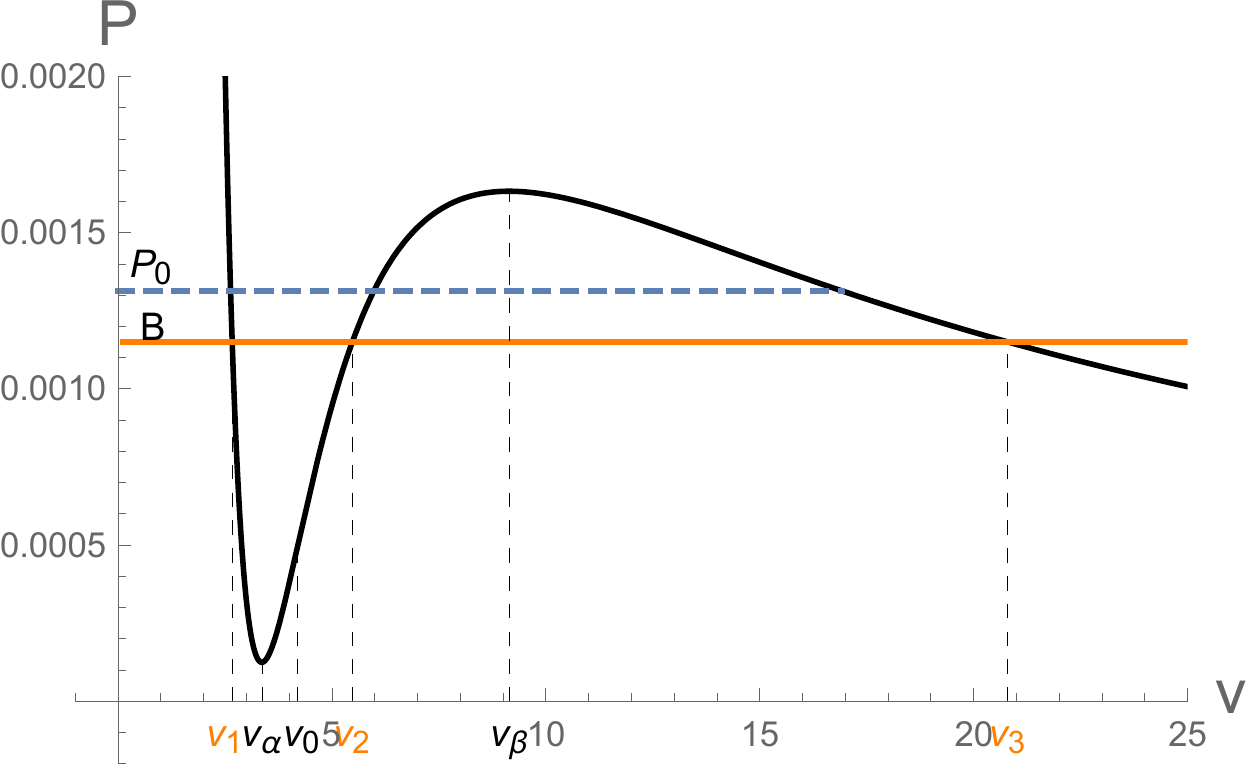}}\quad
\subfigure[]{\includegraphics[width=0.48\textwidth]{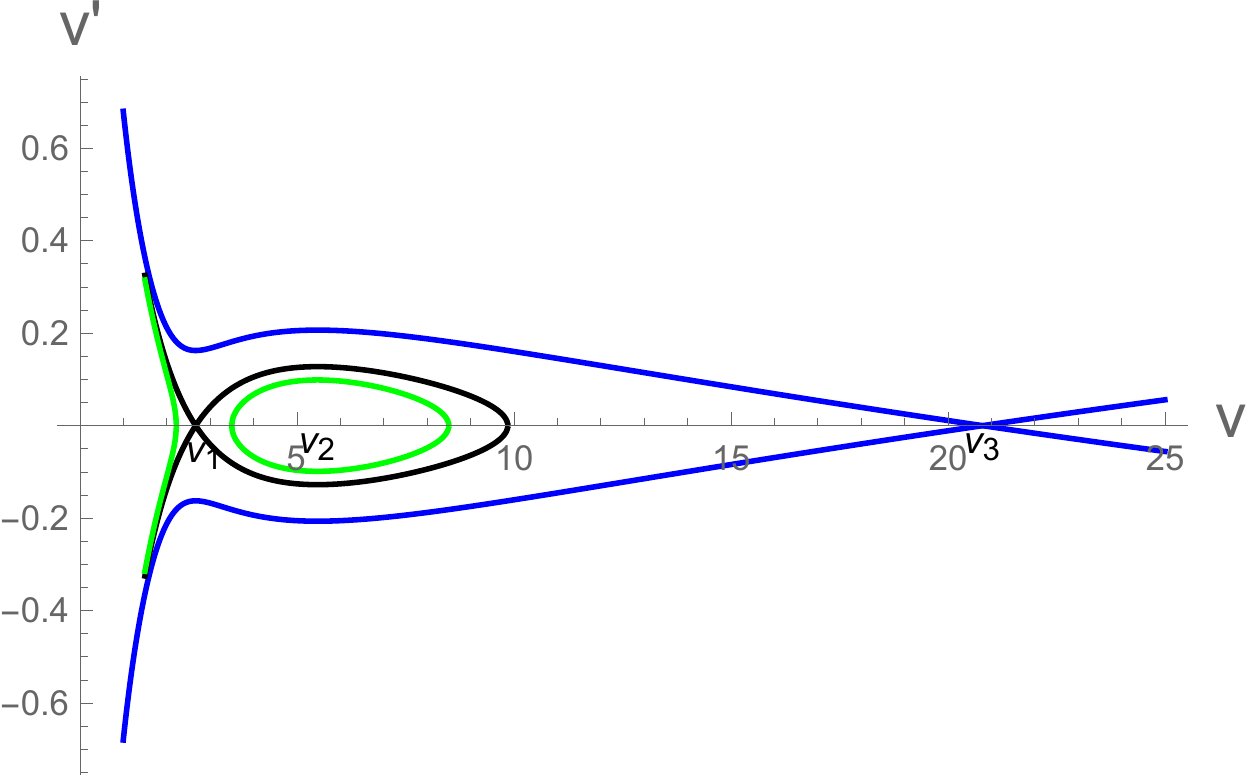}}
\caption{(colour online) Case 2: (a) $B$ and $v_1,v_2,v_3$ in $P-v$ plane; (b) $v-v'$ phase portrait. There is a homoclinic orbit connecting $v_1$ to itself (black one).}
\end{figure}
\begin{figure}[!htbp]
\centering
\subfigure[]{\includegraphics[width=0.48\textwidth]{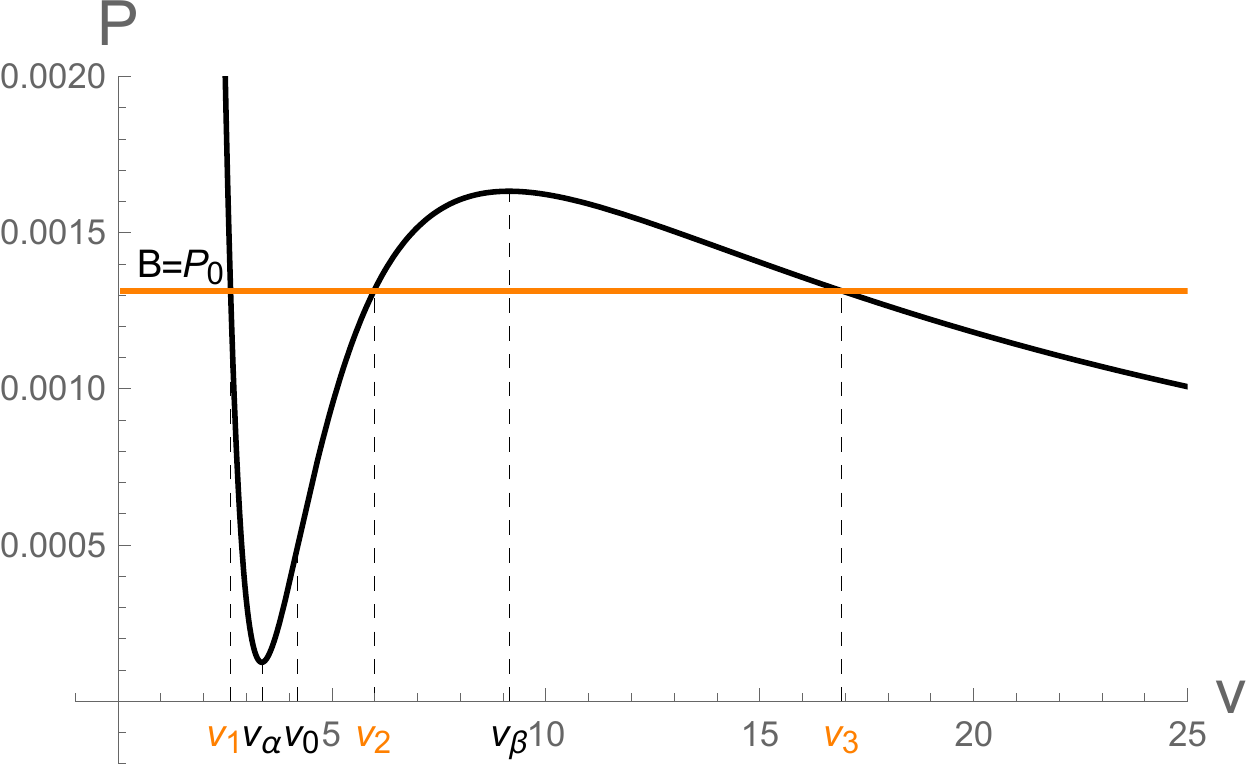}}\quad
\subfigure[]{\includegraphics[width=0.48\textwidth]{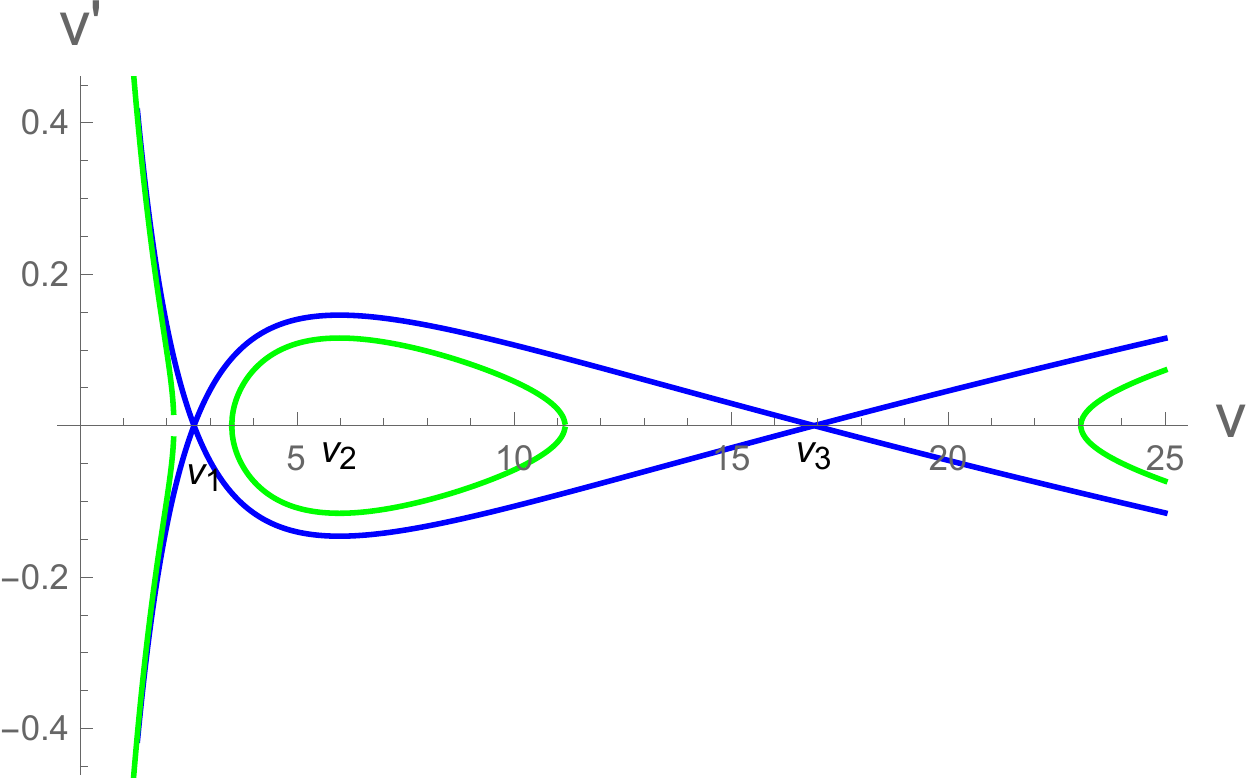}}
\caption{(colour online) Case 3: (a) $B$ and $v_1,v_2,v_3$ in $P-v$ plane; (b) $v-v'$ phase portrait. There is a heteroclinic orbit connecting $v_1$ to $v_3$ (blue one).}
\end{figure}
From the figures, one can see that in case 1 there is a homoclinic orbit $V_1(x)$ connecting the metastable state $v_3$ to itself, i.e., $V_1(x)\rightarrow v_3$ as $x\rightarrow \pm \infty$, while $V'_1(x)\rightarrow 0$ as $x \rightarrow \pm \infty$; The case 2 has a homoclinic orbit connecting $v_1$ to itself; In case 3, there is a heteroclinic orbit $V_3(x)$ connecting $v_1$ and $v_3$, i.e., $V_3(x)\rightarrow v_1 $ as $x\rightarrow -\infty$, $V_3(x)\rightarrow v_3$ as $x\rightarrow+\infty$, while $V'_3(x)\rightarrow 0$ as $x\rightarrow \pm\infty$.

Now let us consider a spatially periodic thermal perturbation as 
\be{
T=T_0 +\epsilon \cos{p x},
}\ee
Then equation (\ref{MainEq2}) becomes
\be{
Av''+P(v,T_0)+ \frac{\epsilon \cos{px}}{v}=B,\label{MainEq3}
}\ee
which again takes the form as Eq.(\ref{GeneralEq}) with $\mathbf{z} \equiv [v, v']^T$ now. So the Melnikov function in three cases can be expressed as
\be{
M(x_0)=\int^{\infty}_{-\infty} -\frac{v'_0(x-x_0)\cos{px}}{v_0(x-x_0)}dx,
}\ee
where $v_0(x)$ and $v'_0(x)$ denoting either the homoclinic or heteroclinic orbit $\mathbf{z}_0(x)=[v_0(x), v'_0(x)]^T$. It is convenient to write this function as
\be{
M(x_0)=-L\cos{px_0}+N\sin{px_0},
}\ee
with
\be{
L=\int^{\infty}_{-\infty}\frac{v'_0(x)\cos{p x}}{v_0(x)}dx,\quad
N=\int^{\infty}_{-\infty}\frac{v'_0(x)\sin{p x}}{v_0(x)}dx.
}\ee

\begin{figure}[!htbp]
	\centering
	\subfigure[~~Case 1]{\includegraphics[width=0.31\textwidth]{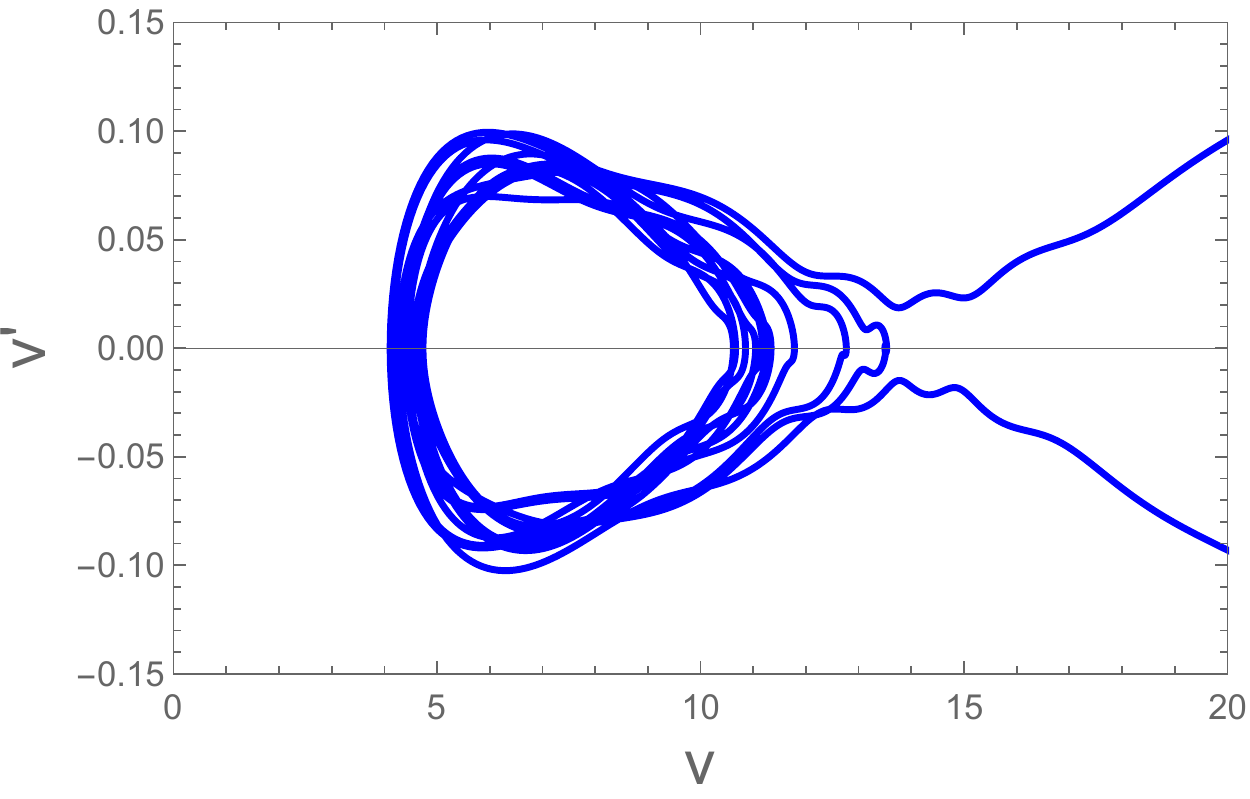}}\quad
	\subfigure[~~Case 2]{\includegraphics[width=0.31\textwidth]{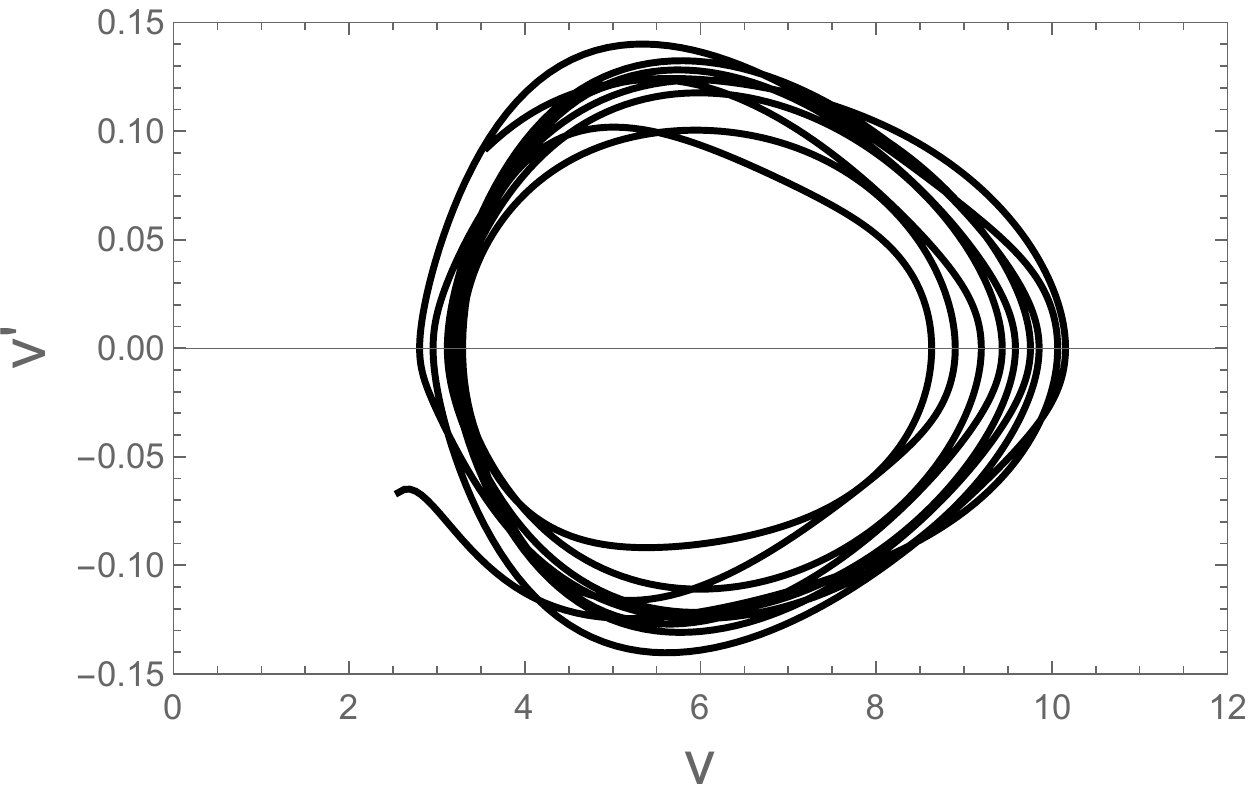}}\quad
	\subfigure[~~Case 3]{\includegraphics[width=0.31\textwidth]{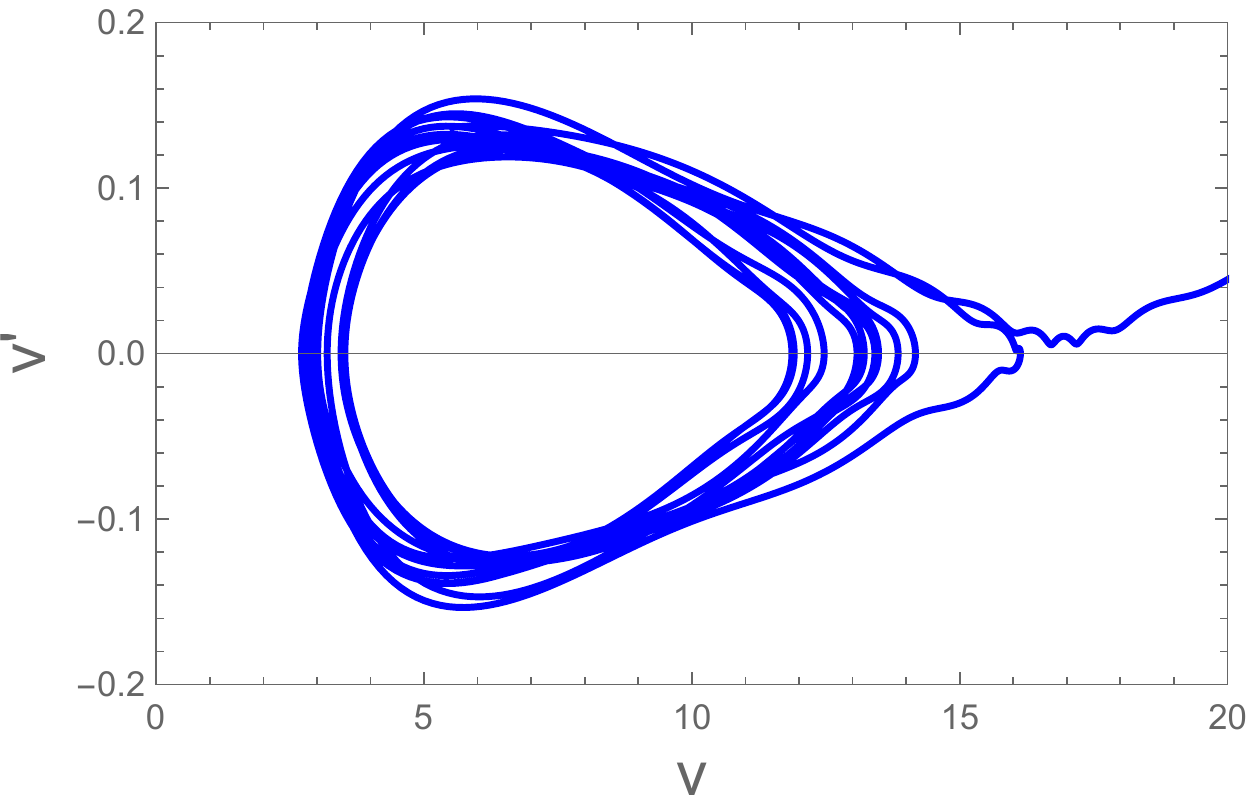}}\quad
	\caption{(colour online) Portrait of the perturbed equation in $v-v'$ phase plane for the three cases. The initial states are the fixed points of the homoclinic orbit or heteroclinic orbit. $\epsilon=0.001, p=0.1$ with other parameters fixed as in Figs. 5-7.}
\end{figure}

Hence, we see that whatever the values of $L$ and $N$ are, $M(x_0)$ always possesses simple zeros, signalling spatial chaos. This result is the same as Refs.~\cite{Slemrod:1985,Chabab:2018lzf,Mahish:2019tgv}. In Fig. 8, the solutions of the perturbed equation (\ref{MainEq3}) for the three cases are plotted in the $v-v'$ plane, with the initial configures chosen to be the homoclinic orbit or heteroclinic orbit. From the figure, one can see that there is indeed spatial chaos under perturbations.

\section{Summary and Discussions}

In extended phase space,BIAdS BH exhibits phase structures resembling that of van der Waals fluid when $b Q>1/2$ in four-dimensional spacetime. We study dynamics of its configuration under temporal thermal perturbations which is assumed to be time periodic. Initially, the system is in the unstable spinodal region, and allows existence of homoclinic orbit in phase space. By applying the Melnikov method, we show that there exists a critical amplitude of the perturbations $\gamma_c$ such that chaos emerges for $\gamma >\gamma_c$,which means that the time evolution of the configuration of the system is unpredictable practically under perturbations. Dependence of $\gamma_c$ on the Born-Infeld parameter $b$ and the black hole charge $Q$ is also studied, and the results show that larger $b$ or $Q$ makes the onset of chaos easier. Moreover, we also consider the spatial thermal perturbations which is periodic in space. And it is found that spatial chaos always exists for whatever the perturbation amplitude is.

In this work, we only consider four-dimensional case with $b Q>1/2$ for simplicity. Extensions to higher dimension and other BHs are straightforward. It is also interesting to see if such phenomena exist in cases where RPT will occur. We leave these to further investigations.

\section*{Acknowledgement}

This work is supported by National Natural Science Foundation of China (No. 11605155).

\end{document}